\documentclass[10pt,twocolumn,letterpaper, fleqn]{article}

\usepackage[pagenumbers]{cvpr} % To force page numbers, e.g. for an arXiv version

\usepackage{graphicx}
\usepackage{siunitx}
\usepackage{amsmath}
\usepackage{amssymb}
\usepackage{booktabs}
\usepackage{paralist}
\usepackage{diffcoeff}
\usepackage[pagebackref,breaklinks,colorlinks]{hyperref}

\usepackage[capitalize]{cleveref}
\crefname{section}{Sec.}{Secs.}
\Crefname{section}{Section}{Sections}
\Crefname{table}{Table}{Tables} 
\crefname{table}{Tab.}{Tabs.}

\def\cvprPaperID{*****} % *** Enter the CVPR Paper ID here
\def\confName{AI535}
\def\confYear{S2022}

\begin{document}

%\title{Deep Learning-Based Reduced Order Modeling of Nonlinear Time-Dependent PDEs}

\title{Autoencoder-based Dimensionality Reduction for Accelerating the Solution of Nonlinear Time-Dependent PDEs: Transport in Porous Media with Reactions\vspace{+2em}}

\author{Diba Behnoudfar\\
Oregon State University\\
{\tt\small \{behnoudd\}@oregonstate.edu}
}
\maketitle

%%%%%%%%% ABSTRACT
\begin{abstract}
   Physics-based models often involve large systems of parametrized partial differential equations, where design parameters control various properties. However, high-fidelity simulations of such systems on large domains or with high grid resolution can be computationally expensive, for the accurate evaluation of a large number of parameters. Reduced-order modeling has emerged as a solution to reduce the dimensionality of such problems. This work focuses on a nonlinear compression technique using a convolutional autoencoder for accelerating the solution of transport in porous media problems. The model demonstrates successful training, achieving a mean square error (MSE) on the order of \num{1e-3} for the validation data. For an unseen parameter set, the model exhibits mixed performance; it achieves acceptable accuracy for larger time steps but shows lower performance for earlier times. This issue could potentially be resolved by fine-tuning the network architecture.
   
\end{abstract}

%%%%%%%%% BODY TEXT
\section{Introduction}

A wide range of physical phenomena, such as fluid flow, are typically represented by a set of governing equations in the form of parametrized partial differential equations (PDEs) discretized over a computational domain, where a set of design parameters controls properties such as the boundary conditions, the geometry of the computational domain, or physical properties. For applications such as design optimization, a large number of parameters must be evaluated with high accuracy. High-fidelity simulations of large systems are often computationally expensive, requiring large amounts of memory and computational time. In the case of reacting flows with detailed chemical kinetics, the computational cost to carry out fluid dynamics simulations is very high because of the large number of species and reactions that must be considered \cite{d2022automated}.\\
Alternatively, reduced-order modeling can be used to reduce the dimensionality of the problem . Recently this technique has been applied to a range of problems, mostly involving fluid flow, for parametric studies \cite{d2022automated, kadeethum2021framework, kim2022fast, laakmann2021efficient, decaria2020artificial, bukka2021assessment, kadeethum2022non}. The method creates compressed representations using training data from a set of computed simulations of the high-fidelity full-order model that allows for rapid, real-time evaluation of simulations at unseen design parameters. There are linear and non-linear methods for data compression. The linear methods are mostly based on principal component analysis (PCA), a classical technique that involves computing the covariance matrix, performing eigen decomposition, and selecting the principal components based on eigenvalues. The data is then projected onto the selected principal components. Due its linear nature, PCA sometimes can not handle the nonlinearities in the problem. For example in the reacting flow problems, the reaction source term makes the problem highly nonlinear. \\
For nonlinear compression, an option is using a deep convolutional autoencoder. Convolutional autoencoders perform well at learning data that are spatially distributed, including the solutions to partial differential equations (PDEs) discretized over a computational domain \cite{hasegawa2020machine}. Autoencoder neural networks consist of two parts: an encoder, which maps high-dimensional inputs to a low-dimensional code, and a decoder, which maps the low-dimensional code to an approximation of the high-dimensional input. The reduced order model can then be solved for any new input parameter set by seeking an approximated solution in the reduced space. This work proposes to use the concept of convolutional autoencoder for the problem of multi-species porous material combustion. Understanding this phenomenon is important for controlling many natural and engineered systems such as wildfires. The eventual goal is to develop a framework for accelerated parametric studies involving this phenomenon.

\section{Methodology}
\label{sec:intro}

This work focuses on solving the problem of one-dimensional combustion inside porous media, composed of separate solid and gas phases. The governing partial differential equations (PDEs) describing the problem are, 

\begin{gather}
\diffp {}t \big(\Bar{\rho_{s}}\Bar{c_{s}}  T\big)  = \diffp {}x\left( \Bar{k_s} \diffp{T}x\right) + \sum_{k=1}^{N_R}\dot\omega_{k}''' \Delta h_{k} \\ 
\diffp {}t \big(\Bar{\rho_{s}} ) = -\sum_{j=1}^{N_S}\dot\omega_{j}'''  \\
\diffp {}t \big(\Bar{\rho_{g}}\phi\big) + \diffp {}x \left(\dot m''\right) =\sum\dot\omega_{j}''' \\ 
\diffp {}t \left(\Bar{\rho_{g}}\phi Y_j\right) + \diffp {}x \left(\dot m''  Y_j\right) =\diffp {}x \left(\Bar{\rho_{g}}\phi D \diffp {Y_j}z\right)+ \dot\omega_{j}'''  \\
 \dot\omega_{i}'''\rvert_A = Z_A T^n \exp{\left(\frac{E_A}{RT}\right)} m_i''' \\
\dot m'' = -\frac{K}{\nu}\left(\diffp {P}x \right) \label{eq:13} \\
P = \frac{\bar\rho_g }{Mw}R T  
\end{gather}

where \textit{T} is temperature; $t$ is time and $x$ is the spacial location; $\dot\omega_{k}'''$ and $\Delta h_{k}$ denote the net destruction rate of reactant \textit{k} and enthalpy of reaction \textit{k}; $\dot\omega_{j}'''$ is the net production rate of gaseous species \textit{j}; $\Bar{\rho_{s}}$, $\Bar{c_{s}}$ and $\Bar{k_s}$ are the mixture-averaged solid density, specific heat capacity, and conductivity; $\Bar{\rho_{g}}$ is the gas density, $Y_j$ is gas phase species mass fraction, $\dot m''$ is mass flux and $D$ is the effective binary diffusion coefficient; $Z_A$ and $E_A$ are pre-exponential factor and activation energy for reaction A and $m_i'''$ is the local mass of solid phase reactant $i$ per unit volume; \textit{K} is permeability, $\nu$ is kinematic viscosity, \textit{P} is pressure and \textit{Mw} denotes average molecular weight. %Appendix A provides more details on the physics-based model.

\textit{T} and $Y_j$  are the primary unknown variables which are functions of the initial compositions of the porous material constituents; therfore these compositions are the parameters of the model. The above-mentioned system of PDEs can be solved numerically by discretizing the domain and applying the above equations to each computational grid cell.% (see Appendix for more details).

\subsection{Reduced order model} 
To accelerate the solution process, this works employs a reduced-order modeling strategy based on convolutional neural networks similar to the method proposed in \cite{kadeethum2022non}. The model construction phase is comprised of three consecutive main steps. \\

\begin{compactenum}[1)]
    \item \textbf{Training Data Generation.} To explore the parametric dependence of the system, the PDEs are solved numerically for a range of parameter sets \(C\). Each parameter set \(C^{(i)}\), contains the initial compositions of the material constituents (with a count of \(P\)) going through combustion reactions:\\

    $C^i = \{C^i_1 \,,C^i_2, \, \dots , C^i_P \}$ , \quad $i = 1, 2, \dots , M$.\\

    The choice of $M$ and the sampling procedure is typically user- and problem dependent. For each parameter set \(C^{(i)}\), the numerical solver outputs a time series of the primary variables' snapshots; each snapshot is a vector containing the solution at each grid point (with a total of $N_G$ points) at time step $t$:\\
    
    $T^t(C^i) = [T_1^t, \, \dots \,, T^t_{N_G}]$. \\
    %\\
    %$Y_j^t(C^i) = [Y_{j,1}^t, \, \dots \,, Y^t_{j, N_G}]$. \\
    %T = 
    %\begin{bmatrix}
    % $T_{1,1}^i$ & \dots & 4\\
    %\vdots & \dots & \vdots\\
    %3 & \dots & 5
    %\end{bmatrix}

    We will focus on expressions and operations carried out on the temperature field ($T$) for the rest of this work. Based on the parameter set cardinality $M$ and the number of time steps ($N_t$), a total of $N_tM$ training examples are generated to be employed in the next steps. The primary variables are normalized to be in the range [0 1] as follows\\
    \begin{equation}
    \frac{T(.;t,C^i)-\text{min}(T)}{\text{max}(T)-\text{min}(T)}.
    \end{equation}

    \begin{table}[t]
  \centering
  \caption{Autoencoder architecture }
  \begin{tabular}{@{}l l l@{}}
    \toprule
    Block & Input size & Output size \\
    \midrule
    1st convolutional layer & [B 1 16] & [B 32 16]\\
    1st contracting block & [B 32 16] & [B 64 8] \\
    2nd contracting block & [B 64 8] & [B 128 4]\\
    3rd contracting block & [B 128 4] & [B 256 2]\\
    4th contracting block & [B 256 2] & [B 512 1]\\
    1st bottleneck & [B 512] & [B z]\\
    2nd bottleneck & [B z] & [B 512] \\
    1st extracting block & [B 512 1] & [B 256 2] \\
    2nd extracting block & [B 256 2] & [B 128 4] \\
    3rd extracting block & [B 128 4] & [B 64 8] \\
    4th extracting block & [B 64 8] & [B 32 16] \\
    2nd convolutional layer & [B 32 16] & [B 1 16]\\
    \bottomrule
  \end{tabular}
  \label{tab:ae}
\end{table}

    \item \textbf{Data Compression.} The first part of the algorithm compresses the information provided by the snapshots using a deep convolutional auto encoder (AE). The AE is composed of three main components: encoder, bottleneck and the decoder. The encoder produces non-linear manifolds, $z^T \, (z_1^T, \, \dots \, , z_Q^T)$ and $z^Y \, (z_1^Y, \, \dots \, , z_Q^Y)$ of temperature and mass fraction, respectively:\\
    
    $z^T(t,C) = \text{encoder}(T(t,C))$\\
    %$z^Y(t,C) = \text{encoder}(Y(t,C))$.\\
    
    $Q$ represents the subspace size or degree of compression; the goal is to achive $Q << N_G$. The $z^T$ lies within the bottleneck layer. The decoder then, reconstructs $\hat{T}$ and $\hat{Y}$ given the compressed data:\\

    $\hat{T}(t,C) = \text{decoder}(z^T(t,C))$\\
    %$\hat{Y}(t,C) = \text{decoder}(z^Y(t,C))$.\\
    
\begin{table}[t]
  \centering
  \caption{Example case information}
  \begin{tabular}{@{}lc@{}}
    \toprule
     & Value \\
    \midrule
    Number of parameter sets (M) & 6 \\
    Number of time steps ($N_t$) & 1001 \\
    $t$ range  & [0 10]\\
    Total dataset size ($MN_t$) & 6006\\
    Training set size & 4004 \\
    Validation set size & 1001\\
    Test set size & 1001\\
    Grid size ($N_G$) & 16 \\
    Reduced subspace dimension ($Q$) & 4, 6, 8 \\
    \bottomrule
  \end{tabular}
  \label{tab:info}
\end{table}

\begin{table*}[htbp]
    \centering
    \caption{Instances of the parameter set $C^{(i)}$ used in the example case}
    \begin{tabular}[t]{@{}*{7}{c}@{}}
    \toprule
    $C^{(i)}$ & $C^1$ & $C^2$ & $C^3$ & $C^4$ & $C^5$ & $C^6$  \\
       \midrule
        
        $C^{(i)}_1$ & 0.4571 & 0.4386 & 0.4748 & 0.4009 & 0.3911 & 0.5859\\
        $C^{(i)}_2$ & 0.1752 & 0.2257 & 0.3658 & 0.3012 & 0.1657 & 0.2571\\
        $C^{(i)}_3$ & 0.0186 & 0.0888 & 0.0030 & 0.0249 & 0.1076 & 0.0458\\
        $C^{(i)}_4$ & 0.1861 & 0.1366 & 0.0982 & 0.1623 &  & 0.0002\\
        $C^{(i)}_5$ & 0.0652 & 0.0202 & 0.0015 & 0.0008 & 0.2573 & 0.0068\\
        $C^{(i)}_6$ & 0.0121 & 0.0430 & 0.0238 & 0.0146 & 0.0222 & 0.1042\\
        $C^{(i)}_7$ & 0.0844 & 0.0471 & 0.0330 & 0.0952 & 0.0562 & \\
        \bottomrule
    \end{tabular}
    \label{table:params}
\end{table*}

    Similar to the work of Kadeethum et al. \cite{kadeethum2022non}, the encoder uses a contracting block with two convolutions (kernel size = 3, padding = 1) followed by a max pool operation;  this block uses  LeakyReLU with a negative slope of 0.3 as activation function. The bottleneck is composed of two linear layers which map the resulting tensor of the form [\# channels, 1] to [$Q$, 1] and vice versa. The decoder inverts these operations using an expanding block with a convolution layer for upsampling (kernel size=2, stride=2) followed by two other convolutions. ReLU is used as the activation function in this block. The model uses batch normalization before each activation step. Table~\ref{tab:ae} presents the details of the AE structure. ADAM algorithm (with a batch size of 32) is used for minimizing the Mean Square Error (MSE) loss function:\\
    \begin{equation}
        MSE^T = \frac{1}{MN_t}\sum_{i=1}^{M}\sum_{k=0}^{N_t}|\hat{T}(t^k,C^i) - T(t^k,C^i)|^2.
    \end{equation}

    \item \textbf{Reduced Subspace Predictor.} The ultimate application of the model developed here is to predict $z^T$ and $z^Y$ given an arbitrary $(t, C)$, and then reconstruct the numerical solution using the reduced representation. To achieve this, the second part of the algorithm trains a deep neural network that maps $(t, C)$ to $z$. The network is made of five fully-connected linear layers with seven neurons and a $tanh$ activation function, similar to the work of Kadeethum et al. \cite{kadeethum2022non}. The data available for this task are the pairs of $(t, C)$ in the training set and the resulting reduced representation vectors, $z$, which are all normalized to [0 1]. As in the previous step, ADAM algorithm is used to minimize the following loss function:\\
    \begin{equation}
        MSE^{z^T} = \frac{1}{MN_t}\sum_{i=1}^{M}\sum_{k=0}^{N_t}|\hat{z}^T(t^k,C^i) - z^T(t^k,C^i)|^2.
    \end{equation}

\end{compactenum}

\noindent During the inference phase, we query the reduced subspace predictor for a desired time and parameter set, and then reconstruct the primary variables using the decoder.

\begin{figure}
  \centering
  %\begin{subfigure}[t]{0.68\linewidth}
    \captionsetup{width=1\linewidth}
    %\fbox{\rule{0pt}{2in} \rule{.9\linewidth}{0pt}}
    \includegraphics[width=1\linewidth]{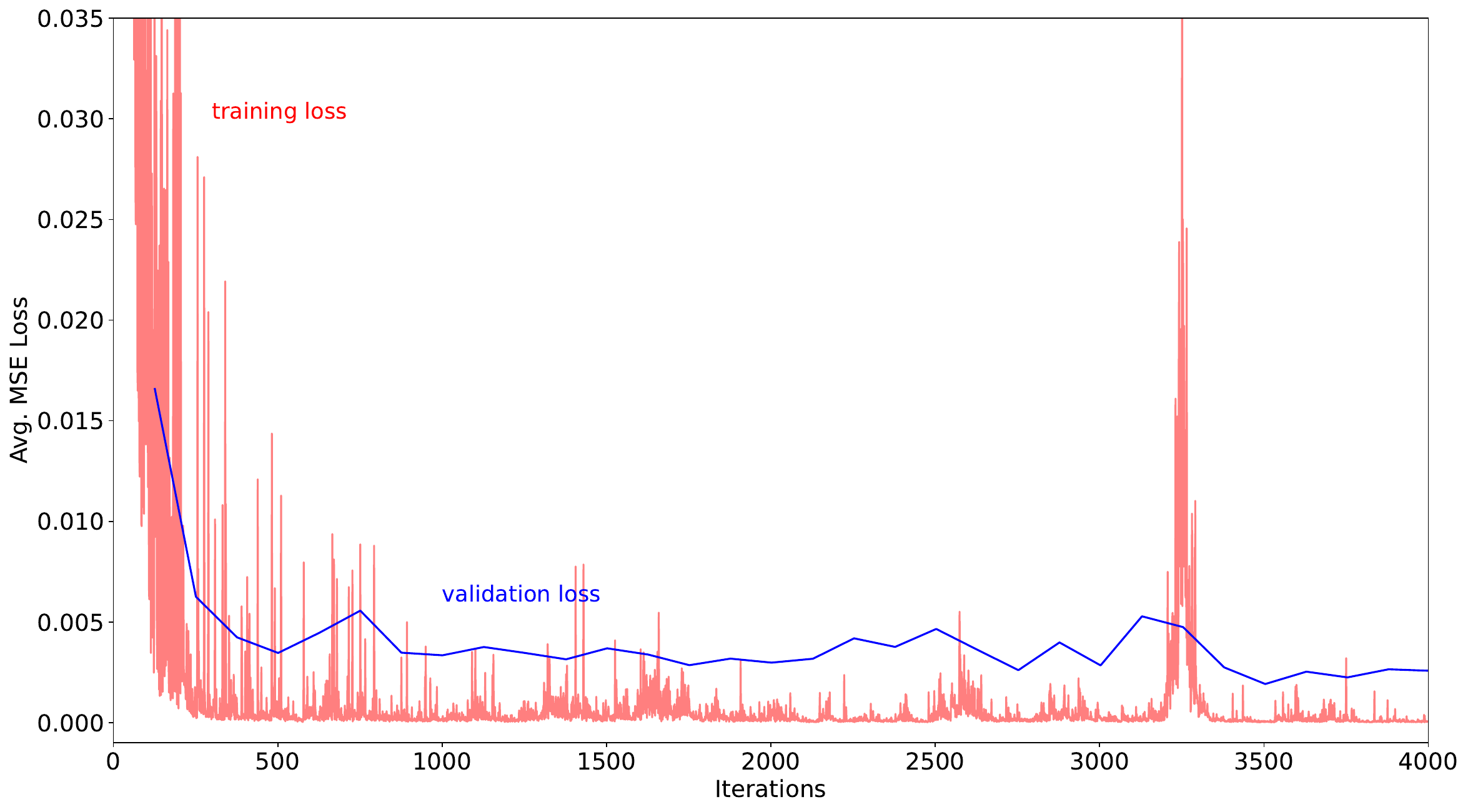}
 
  \caption{Average MSE loss for autoencoder (Q = 4) applied on training and validation sets}
  \label{fig:loss}
\end{figure}

\begin{figure}[t]
    \centering
    \includegraphics[width=1\linewidth]{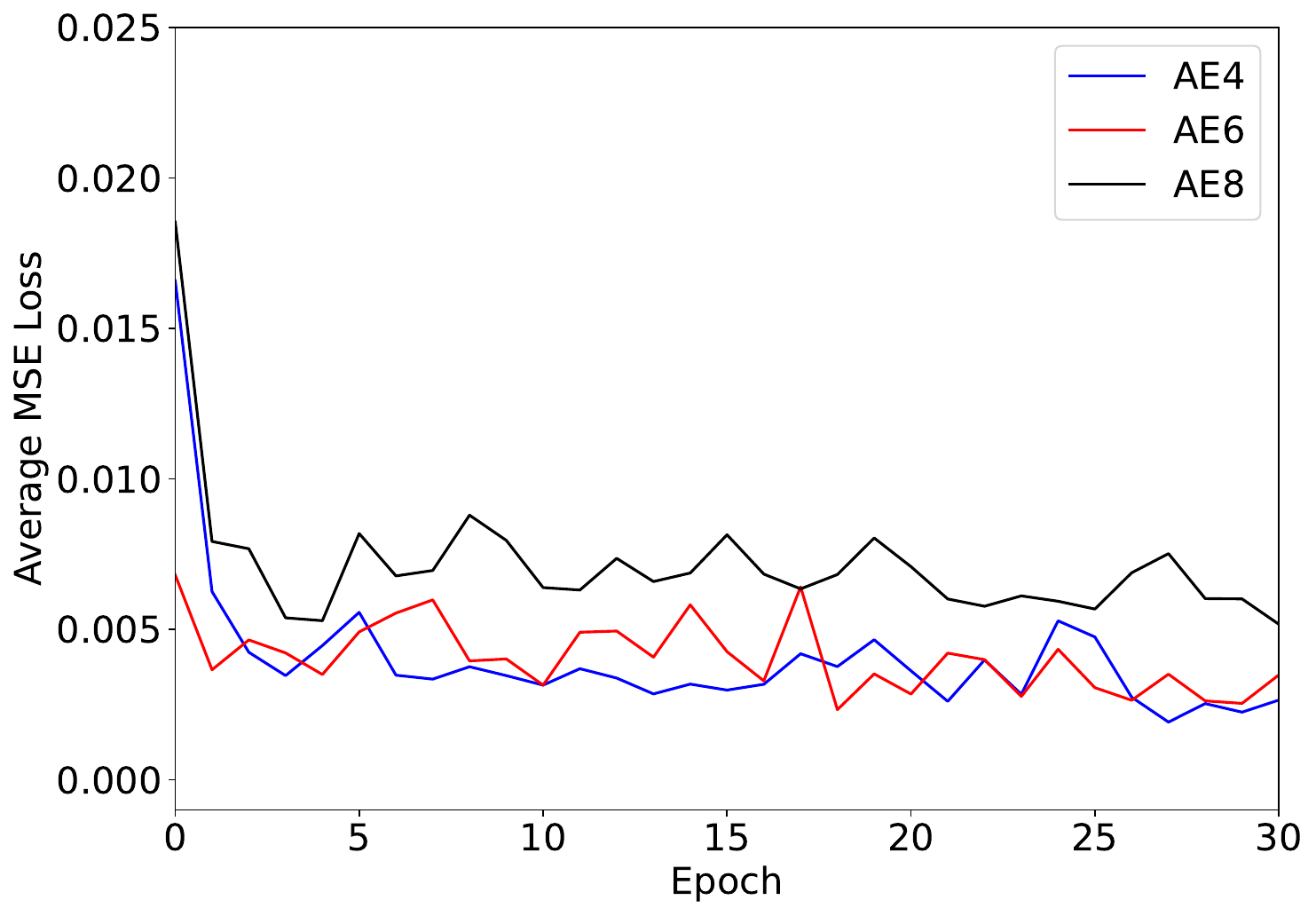}
    \caption{Validation loss of autoencoder with reduced subspace size of 4 (AE4), 6 (AE6), and 8 (AE8)}
    \label{fig:vals}
\end{figure}

\begin{figure}[t]
    \centering
    \includegraphics[width=1\linewidth]{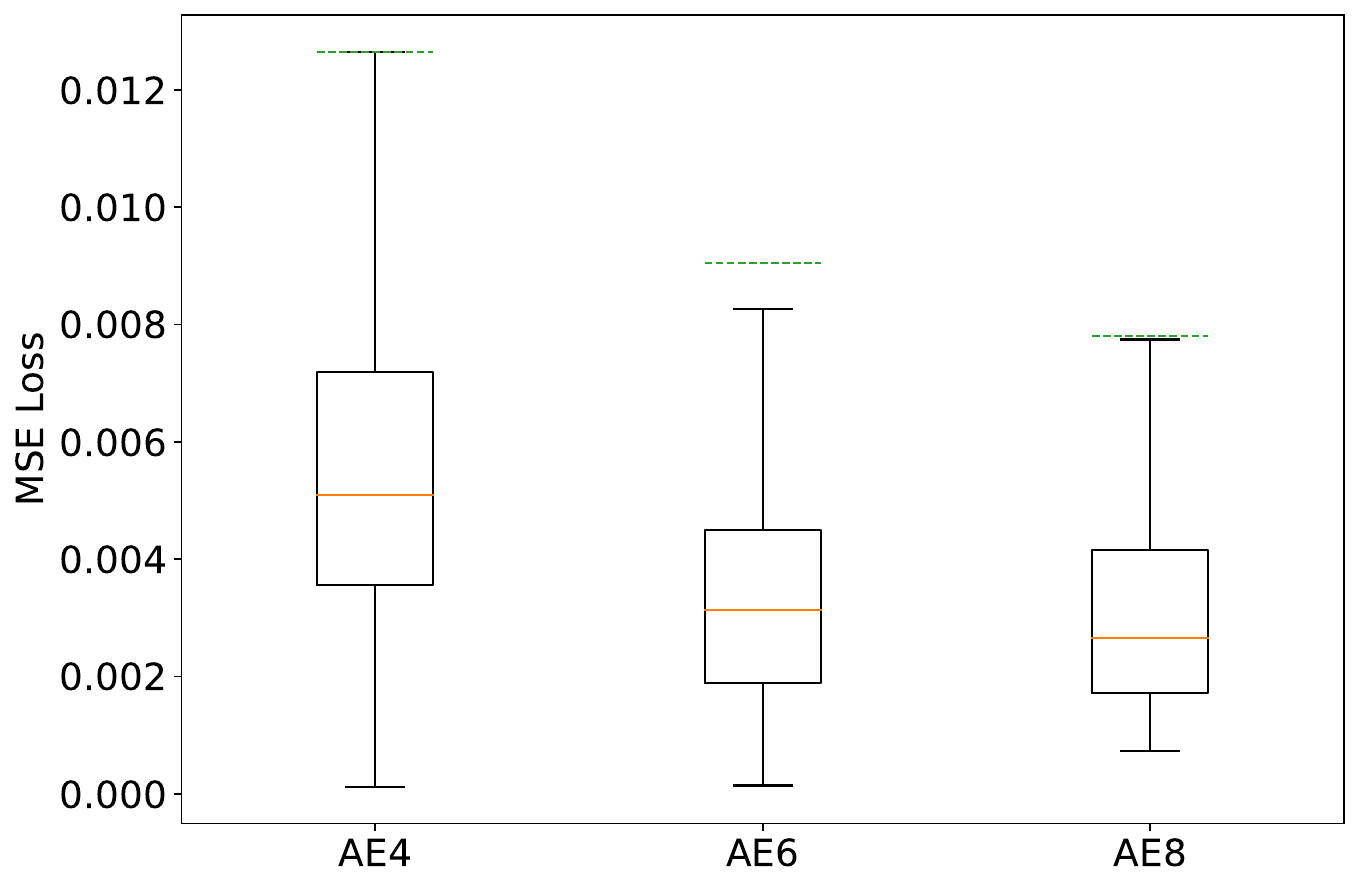}
    \caption{MSE loss distribution for the test data. The box extends from the lower to upper quartile values of the data; the orange line indicates the median and green dotted line the mean; whiskers indicate $1.5 \times \text{inter-quartile range}$}
    \label{fig:box}
\end{figure}

\begin{figure}
  \centering
  \begin{subfigure}[t]{1\linewidth}
    \centering\captionsetup{width=1\linewidth}
    %\fbox{\rule{0pt}{2in} \rule{.9\linewidth}{0pt}}
    \includegraphics[width=1\linewidth]{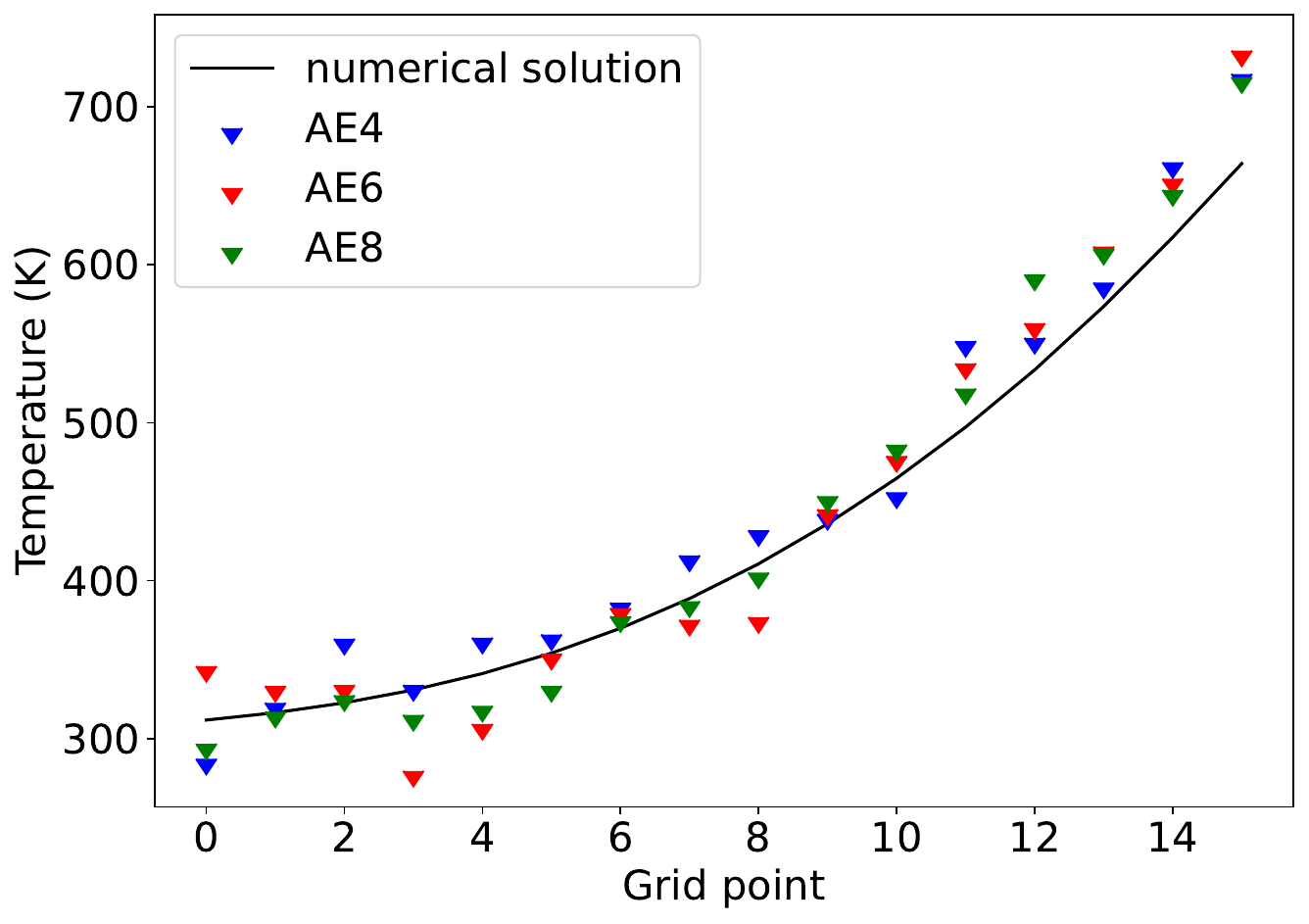}
    \caption{$t = 5$}
    \label{}
  \end{subfigure}
  \hfill
  \begin{subfigure}{1\linewidth}
    %\fbox{\rule{0pt}{2in} \rule{.9\linewidth}{0pt}}
     \includegraphics[width=1\linewidth]{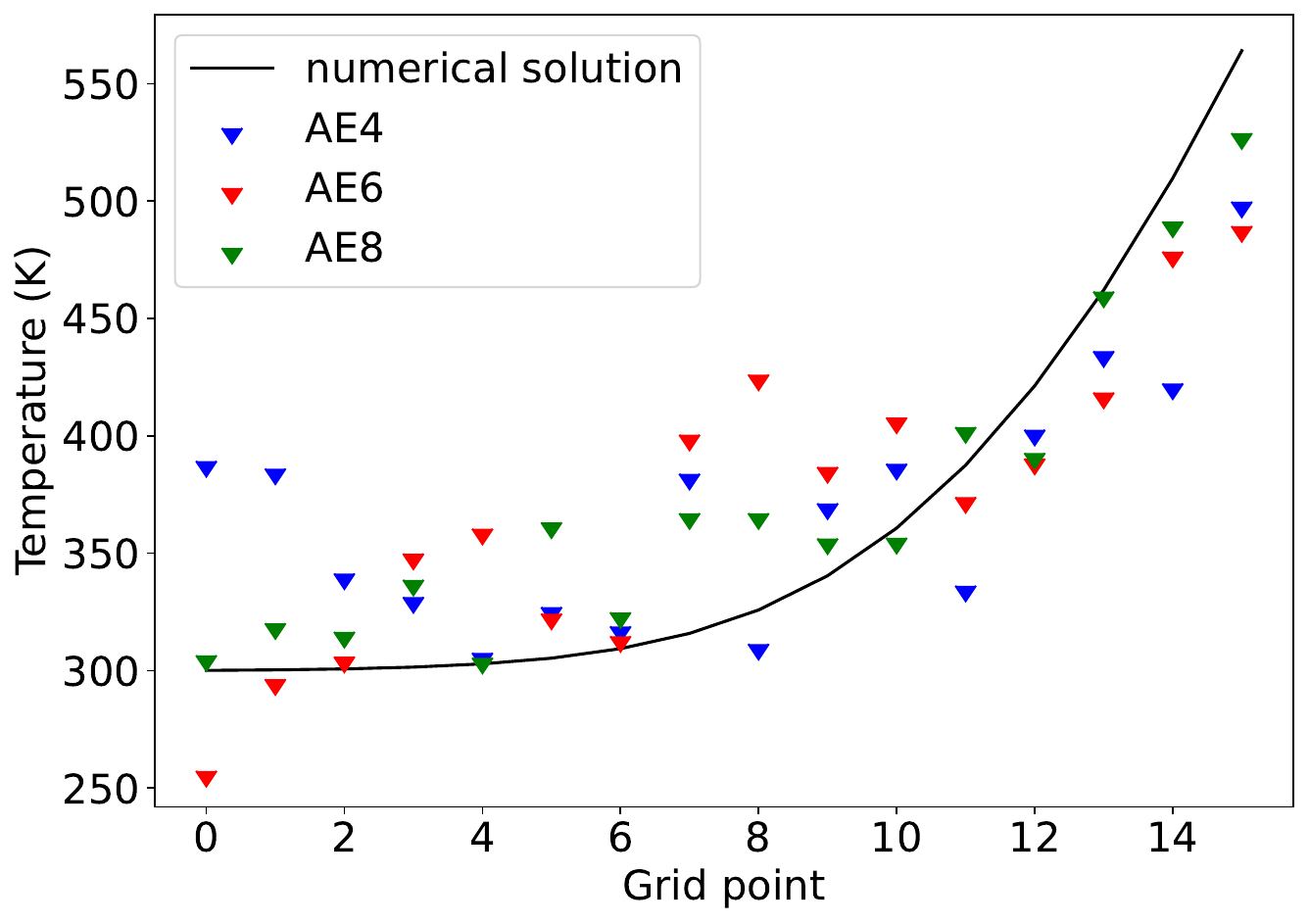}
    \centering\caption{$t = 2$}
    \label{}
  \end{subfigure}
  \caption{Comparison of model results with numerical solution at two different times for the test data}
  \label{fig:temps}
\end{figure}

\section{Results}
This section examines the performance of the proposed reduced order model applied on an example case with the information provided in Table~\ref{tab:info}. We first look at the training process for the auto encoder. The parameter sets in Table~\ref{table:params}, except $C^4$, are used for generating the training and validation data. We keep the data associated with parameter set $C^4$ for testing. The training loss diminishes rather quickly (with some initial oscillations) and reaches a value in the order of \num{1e-5} (Figure~\ref{fig:loss}). The validation loss follows the same trend, however, its value does not reach below \num{1e-3}, which could be an indication that the model is overfitting to the training data. Figure~\ref{fig:vals} compares the validation losses of auto encoders with different reduced subspace dimensions ($Q$). Interestingly, the model with smallest $Q$ (or highest degree of compression), performs best on the validation data and the performance decreases with increasing the $z$ dimension. This behavior implies that using only $Q = 4$, the auto-encoder could capture most of the information. Figure~\ref{fig:box} shows the performance of the model on the test data using different $z$ dimensions ($Q$). We observe that the model accuracy on the validation data transfers well to the majority of the test data, however, with an opposite trend as in Figure~\ref{fig:vals}; both the mean and variation of the error decreases with increasing the $z$ dimension, but this does not guarantee a better model. 

As seen in Figure~\ref{fig:temps}, the local performance of the model varies with time parameter. All three models ($Q = 4,6,8$) perform poorly at early times with $Q = 4$ having the worst performance but it improves at larger $t$ values. Clearly, the large error at early times shifts the average loss to a larger value which explains the trend seen in Figure~\ref{fig:box}.

\section{Discussion}

As reported above, the model developed here struggles with correctly predicting the temperature at early time steps. What characterizes the data from that time duration is the regions with small spatial variation. When a region of an input data has very little variation or subtle changes, the convolutional filters may struggle to capture meaningful features and produce less discriminative representations. Convolutional filters have a limited receptive field. If the region of small variation is larger than the receptive field, the network may not capture the necessary details to differentiate it from other regions. To address this issue several techniques might be helpful including: dilated convolutions, attention mechanisms, data augmentation, and multi-scale and multi-resolution analysis. Alternatively, combining a sequence model with the autoencoder to handle the temporal variations might be effective. Hasegawa et al \cite{hasegawa2020machine} reported that a framework based on CNN autoencoder and a long short term memory (LSTM) worked well for unsteady 
flows around bluff bodies of various shapes.

On the other hand, the model developed here obviously suffers from lack of sufficient training data and spatial resolution. The size of the computational grid in the example studied here is only sixteen which is probably insufficient for capturing the spatial variations in this data. Also, the parameter sets cardinality (M) is very low considering the number of components in each set. Therefore, a more appropriate sampling procedure seems to be necessary for accurate evaluations.

\section{Conclusion}

This work presents a model for reducing the dimensionality of system of partial differential equations with temporal and spatial variation, based on a deep convolutional autoencoder. The model is successfully trained with a mean square error (MSE) on the order of \num{1e-5} for training data and \num{1e-3} for validation data. The model has mixed performance on the test data with acceptable accuracy for larger time steps and lower accuracy for early times. The issue can probably be addressed by fine tuning the network architecture. Another finding of the study is that small reduced subspace dimensions are as effective (or better in some cases) compared to the larger subspace sizes and are able to capture most of the information.
%

%------------------------------------------------------------------------

%-------------------------------------------------------------------------

%%%%%%%%% REFERENCES
{\small
\bibliographystyle{ieee_fullname}
\bibliography{egbib}
}

\end{document}